\documentclass[preprint,12pt]{elsarticle}
\usepackage{amssymb,graphicx,multirow,hyperref}
\journal{Acta Materialia}

\begin{document}

\begin{frontmatter}

\title{Effect of local structural distortions on magnetostructural transformation in Mn$_3$SnC}

\author[gu]{E. T. Dias}
\author[gu]{K. R. Priolkar\corref{krp}}\ead{krp@unigoa.ac.in}
\cortext[krp]{Corresponding author}

\address[gu]{Department of Physics, Goa Univeristy, Taleigao Plateau, Goa 403206 India}

\author[barc]{A. Das},
\address[barc]{Solid State Physics, Division, Bhabha Atomic Research Centre, Trombay, Mumbai 400085}

\author[eu]{G. Aquilanti}
\address[eu]{Elettra-Sincrotrone Trieste S.C.p.A., s.s. 14, km 163.5 I-34149 Basovizza, Trieste, Italy}

\author[turk]{$\rm\ddot{O}$. $\rm\c{C}$akir}
\address[turk]{Physics Department, Yildiz Technical University, TR-34220 Esenler, Istanbul, Turkey}

\author[duis]{M. Acet}
\address[duis]{Faculty of Physics and CENIDE, Universitat Duisburg-Essen, D-47048 Duisburg, Germany}

\author[tifr]{A. K. Nigam}
\address[tifr]{Tata Institute of Fundamental Research, Dr. Homi Bhabha Road, Colaba, Mumbai 400005, India}

\begin{abstract}
In this paper we attempt to understand the different nature of first order magnetic transformation in Mn$_3$SnC as compared to that in Mn$_3$GaC. The transformation in Mn$_3$SnC is close to room temperature (T$_t \approx$ 280K) and is associated with a large change in magnetic entropy that makes it a suitable candidate for applications in ferroic cooling. Using a combination of x-ray and neutron diffraction and x-ray absorption fine structure (XAFS) spectroscopy we infer that the magnetic ground state consisting of antiferromagnetic and ferromagnetic Mn atoms is due to structural distortions present in Mn$_6$C octahedra.
\end{abstract}

\end{frontmatter}


\section{Introduction}
Antiperovskite compounds that have attracted considerable attention due to significant properties like superconductivity \cite{He2001411,Uehara200776},  giant magnetoresistance \cite{Kamishima200063}, magnetostriction effect \citep{Asano200892} and giant negative thermal expansion (NTE) \cite{Takenaka200587,Takenaka200892,Iikubo2008101,Huang200893,Takenaka200994}. Amongst these is Mn$_3$SnC which exhibits a large magnetic entropy change near room temperature with $ \Delta $S$ _{max} $ values ($\sim $80.69mJ/cm$  ^{3}$K and 133mJ/cm$  ^{3}$ K under a magnetic field of 2T and of 4.8T, respectively) comparable to those observed in contemporary magnetic refrigerant materials \citep{Wang200985}. In spite of its relatively simple cubic structure ({\em Sp. Gr. Pm$\bar 3$m}) with Mn atoms located at the face centers of a cube, Sn atoms at the corners and carbon atom positioned at the body center \cite{Howe19572}, Mn$_{3}$SnC transforms from a room temperature paramagnetic (PM) state to a high volume magnetically ordered state with a complicated spin arrangement consisting of antiferromagnetic (AFM) and ferromagnetic (FM) components via a spontaneous first order transition at 280K \cite{Lorthioir19738, Heritier197712, Li200572, Wang200985}. The non collinearity of Mn spins in the transformed state has been attributed to a novel magnetic structure obtained from neutron diffraction studies. Though equivalent in crystal structure, Mn atoms in the magnetic unit cell ($a\sqrt{2}$, $a\sqrt{2}$, $a$) of Mn$_{3}$SnC generated using a propagation vector $k = [{1\over 2}, {1\over 2}, 0]$ are of two types. Firstly, two of the three Mn atoms present themselves in a square configuration in the (001) plane with a net antiferromagnetic moment of 2.4$ \mu_{B} $ per Mn along with a small FM moment of 0.2$ \pm $0.15$ \mu_{B} $ along the [001] direction. While, the remaining Mn atoms have their spins aligned parallel to each other thus contributing to the FM component with a moment of 0.65$ \pm $0.15$ \mu_{B} $ along the 001 direction \cite{Heritier197712}. In contrast, Mn$_3$GaC, which also undergoes a first order transition at T$\sim160K$ has a collinear AFM structure described by a propagation vector $k = [\frac{1}{2}, \frac{1}{2}, \frac{1}{2}]$ resulting in ferromagnetic [111] planes of Mn atoms alternating in their spin direction along the (111) axis \cite{Fruchart19708}.

\section{Experimental}
To prepare polycrystalline Mn$  _{3}$SnC, starting materials (Mn, Sn and C powders) were separately weighed and intimately mixed in the stoichiometric molar ratio 3:1:1 with about 15\% excess graphite powder added to compensate possible carbon deficiencies during the reaction \cite{Lewis200393}. The mixture was then pressed into a pellet and encapsulated in an evacuated quartz tube before sintering at 1073K for the first 48 hours and at 1150K for the next 120 hours \cite{Wang200985}. After quenching to room temperature, the product was pulverized, mixed and annealed again under the same conditions to obtain a homogeneous sample. Room temperature x-ray diffraction (XRD) studies were carried out using a Rigaku diffractometer with CuK$  \alpha$ radiation to determine the phase formation and purity of the compound formed. Thermal expansion across the first order transition was measured from XRD patterns recorded in the temperature range of 25-300K using BL18B at Photon Factory, Japan. Temperature dependent magnetization measurements in the 5-300K range were performed using Quantum Design SQUID magnetometer in an applied field of 0.01T while resistivity measurements in the same temperature range were accomplished using standard four probe method. Neutron powder diffraction 
patterns as a function of temperature were recorded on the PD2 diffractometer ($ \lambda $= 1.2443{\AA}) at Dhruva reactor, Bhabha Atomic Research Centre, Mumbai, India. Rietveld analysis for all diffraction patterns was performed using the FullProf Suite refinement program \cite{Carvajal1993192}. To understand the local structural changes, if any, in the Mn-C octahedra in  Mn$_{3}$SnC, Extended x-ray absorption fine structure (EXAFS) data at the Mn K edge (6539 eV) was collected in transmission mode at 300K (RT) and 80K (LT) in the range from -200 to 1300 eV with respect to the Mn K edge at the XAFS beamline at Elettra, Trieste \cite{Giuliana2009190}. Both incident (I$  _{0}$) and transmitted (I) intensities were measured simultaneously using an ionization chamber filled with appropriate gases. To restrict the absorption edge jump ($ \Delta\mu $) to an optimum value, the number of layers of Mn$  _{3}$SnC powder coated scotch tape were appropriately adjusted. The edge energy was calibrated using Mn metal foil as standard. To obtain EXAFS ($\chi(k)$) signal, Mn K edge data was reduced following standard procedures in Demeter program \cite{Ravel200512}.

\section{Results and Discussion}
\begin{figure}
\begin{center}
\includegraphics[width=\columnwidth]{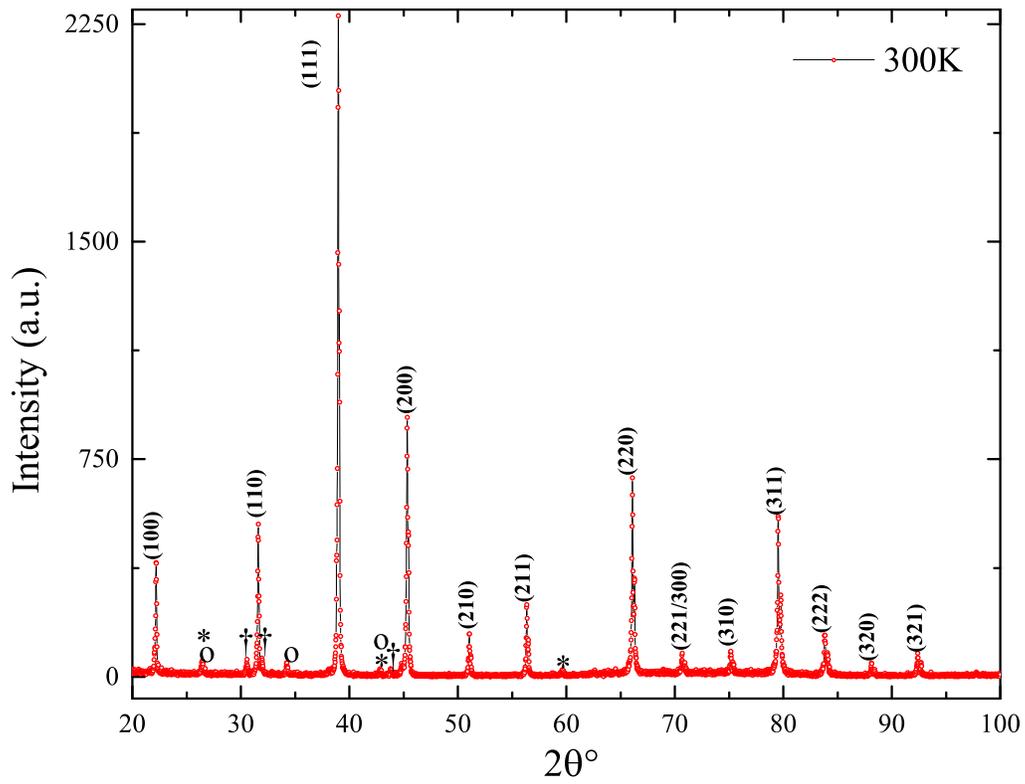}
\caption{X-ray diffraction pattern recorded for Mn$_{3}$SnC at room temperature. $\lq\lq\ast, \dagger, \circ$'' represent minor impurities ($\sim 1\%$) of C, Sn and SnO respectively.}
\label{fig:rtxrd}
\end{center} 
\end{figure}

Room temperature XRD pattern recorded in the angular ($2\theta$) range of 20$ ^{\circ}$  to 100$ ^{\circ}$ in steps of 0.02$ ^{\circ}$ using Cu K$_\alpha$ radiation is presented in Figure \ref{fig:rtxrd}. Rietveld refinement of the pattern shows the formation of cubic Mn$_{3}$SnC with a lattice constant, $a$ = 3.99672(4) \cite{Lorthioir19738} along with minor impurities of C, Sn and SnO.

\begin{figure}
\begin{center}
\includegraphics[width=\columnwidth]{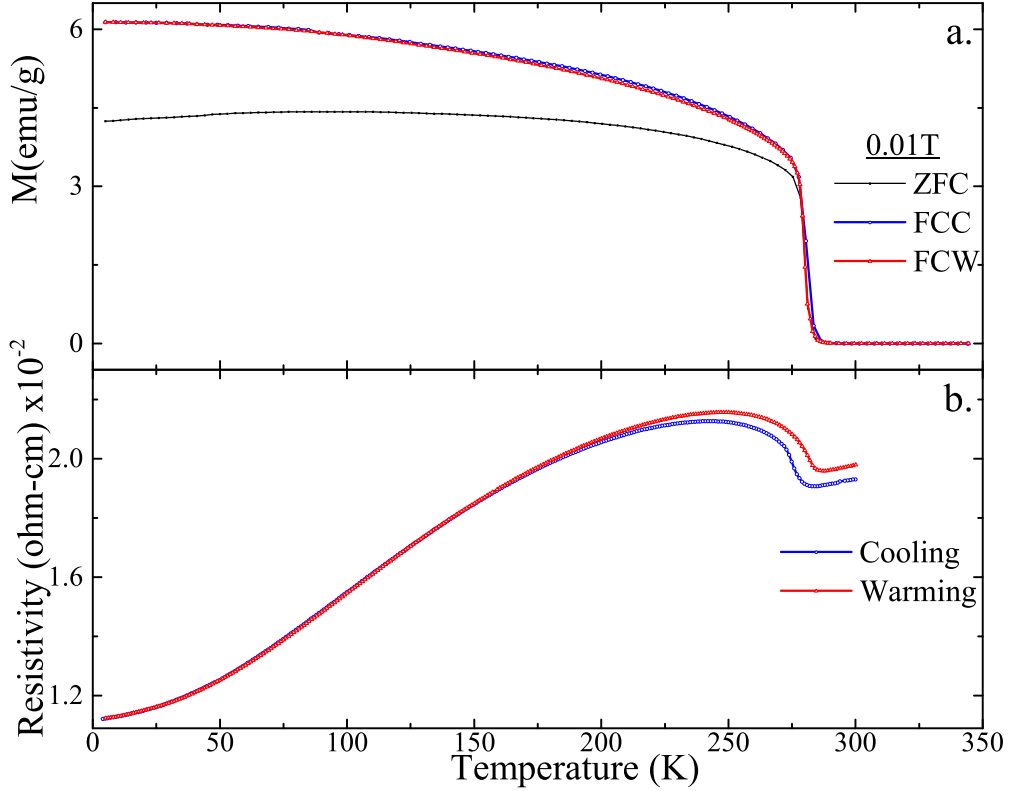}
\caption{Temperature dependence of a. Magnetization under ZFC, FCC and FCW processes in an applied field of 0.01T and b. Electrical resistivity of Mn$_{3}$SnC compound.}
\label{fig:vstemp}
\end{center}
\end{figure}

Magnetization (M) measurements recorded as a function of temperature under zero field cooled (ZFC), field cooled cooling (FCC) and field cooled warming (FCW) conditions at an applied field of 0.01T are shown in Figure \ref{fig:vstemp}a. The M(T) curves exhibit a sharp increase in magnetization at T$ \sim $279 K corresponding to a transition from a high temperature PM state to a low temperature magnetically ordered phase with competing FM and AFM interactions \cite{Heritier197712}. 
Temperature dependent resistivity plots in Figure \ref{fig:vstemp}b measured in zero applied field illustrate that the transport behaviour of Mn$_{3}$SnC is metallic across the 5-350K range \cite{Wang200985} with a sharp discontinuity in resistance around the Curie temperature (T$  _{C}\sim$279K). Existence of thermal hysteresis between the cooling and warming process in both magnetic and transport measurements implies the first order character of the transition at T$_{C}$.

\begin{figure}
\begin{center}
\includegraphics[width=\columnwidth]{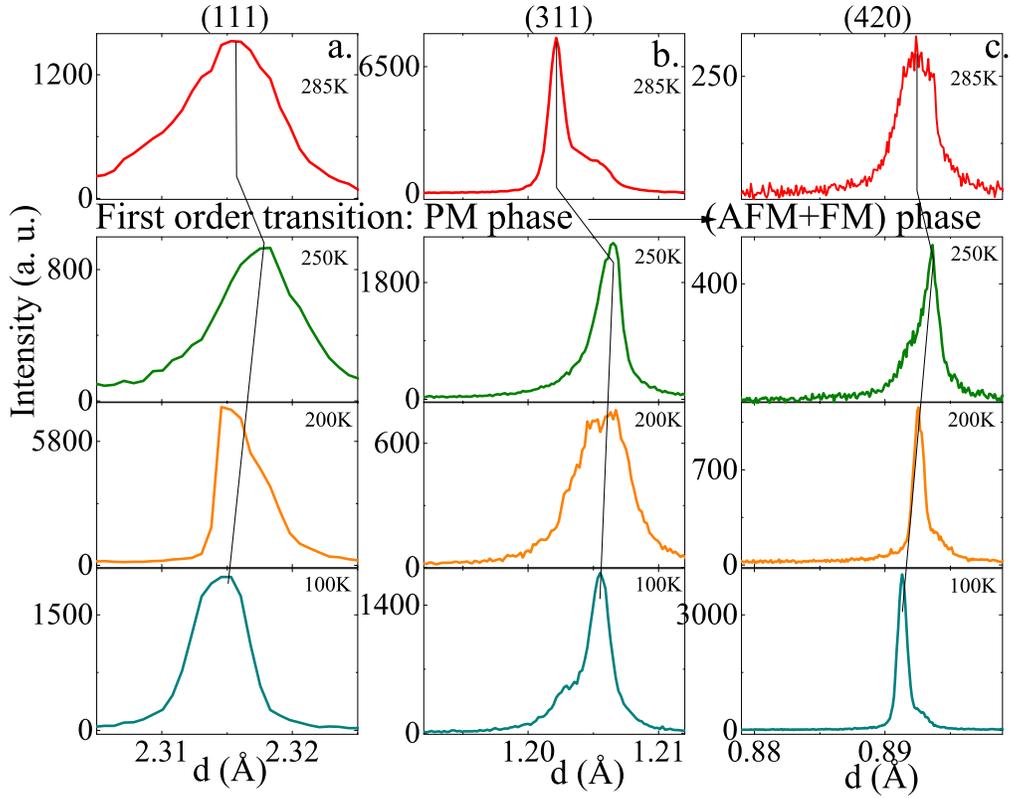}
\caption{Thermal evolution of the (111), (311) and (420) X-ray reflections across the transformation temperature in Mn$_{3}$SnC.}
\label{fig:tempxrd}
\end{center}
\end{figure}

X-ray diffraction intensities in the 25-300K temperature range were collected over a 2$\theta$ range between 8$ ^\circ$ - 45$ ^\circ$  with a step size of 0.005$^\circ$  using an x-ray wavelength $\lambda \sim$ 0.619\AA. Abrupt structural changes that occur around the ordering temperature are highlighted in Figure \ref{fig:tempxrd} via the temperature evolution of the (111), (311) and (420) reflections. One can notice a sudden shift in the position of these reflections to higher values of inter planar distance $\lq d$' between 285K and 250K. This fact coupled with the absence of any additional peak indicates that Mn$_3$SnC undergoes a cubic-cubic volume discontinuous transition at about 279K along with a change in its magnetic state. The calculated change in the lattice constant value at the transition is of the order of 0.1\% which is in agreement with previously reported values \cite{Li200572}.

\begin{figure}
\begin{center}
\includegraphics[width=\columnwidth]{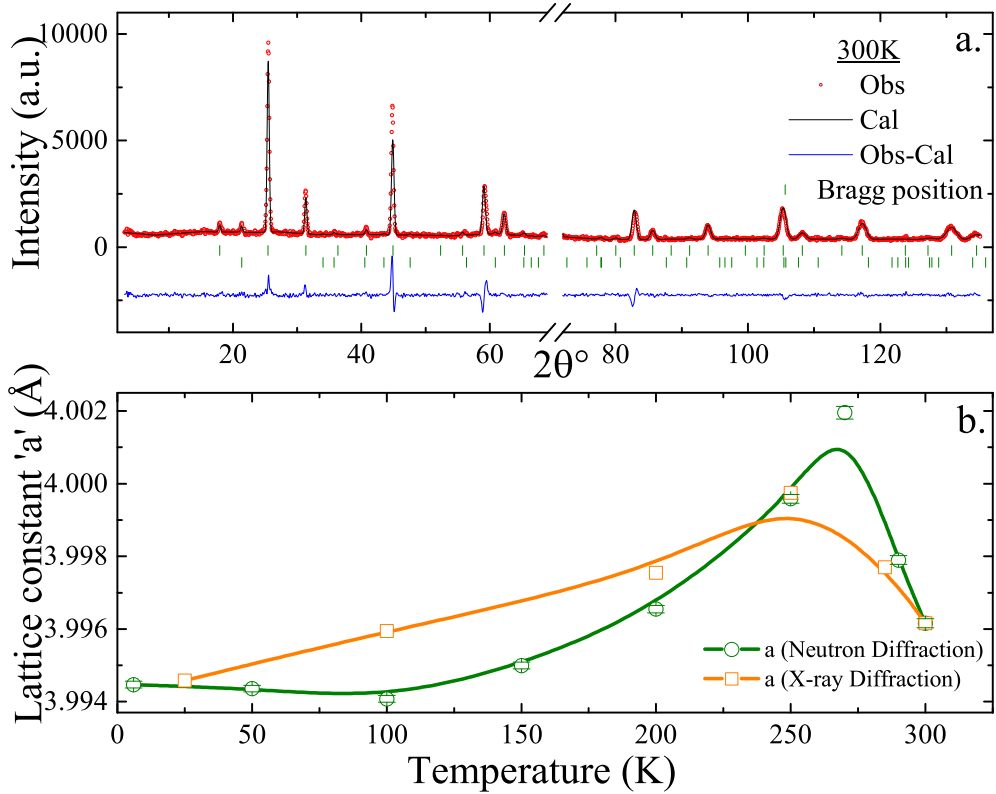}
\caption{a. Rietveld refined neutron diffraction patterns of Mn$_{3}$SnC recorded above the magnetic ordering temperature (T = 300K). b. Variation of refined lattice parameters obtained from neutron diffraction (open circles) and synchrotron x-ray diffraction (open squares) as a function of temperature.}
\label{fig:neutron3}
\end{center}
\end{figure}

As mentioned before the non collinearity of Mn spins observed in Mn$_{3}$SnC below T$ _{C} $ is attributed to a complicated arrangement of FM and AFM spins. To reconfirm magnetic structure below its Curie temperature and to study the thermal evoulution of Mn magnetic moments,  neutron diffraction patterns were recorded at selected temperatures betwen 6K and 300K, in the angular range 3$^{\circ}$ - 135$^{\circ}$. The patterns were Rietveld refined to obtain the crystal and magnetic structure. At 300K, Bragg reflections corresponding only to the nuclear structure (in the space group Pm$\bar 3$m) and impurity phase C are present as can be seen in Figure \ref{fig:neutron3}a. The refined value of lattice constant obtained from this fitting ($a$ = 3.9961(1)\AA) is in good agreement with that obtained from x-ray diffraction. Temperature dependance of lattice constant values obtained from synchrotron x-ray diffraction and neutron diffraction are shown in Figure \ref{fig:neutron3}b. Both the results, obtained from synchrotron x-ray and neutron diffraction, are in fair agreement with each other.

\begin{figure}[h]
\begin{center}
\includegraphics[width=\columnwidth]{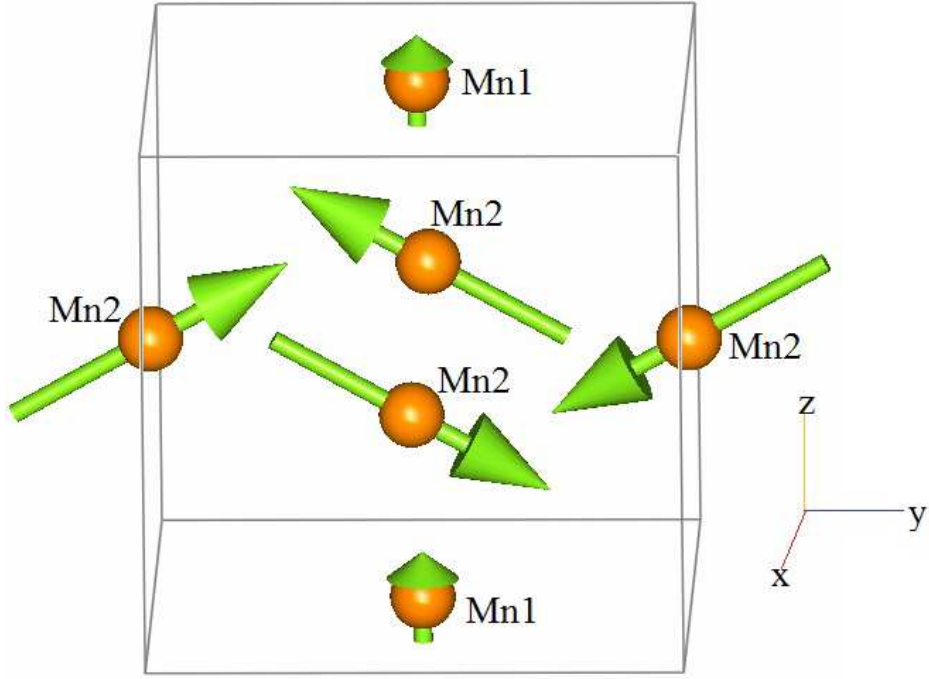}
\caption{Magnetic spin alignment of Mn atoms in nuclear unit cell of Mn$_3$SnC at low temperature as obtained from neutron diffraction.}
\label{fig:neutron5}
\end{center}
\end{figure}

\begin{figure}
\begin{center}
\includegraphics[width=\columnwidth]{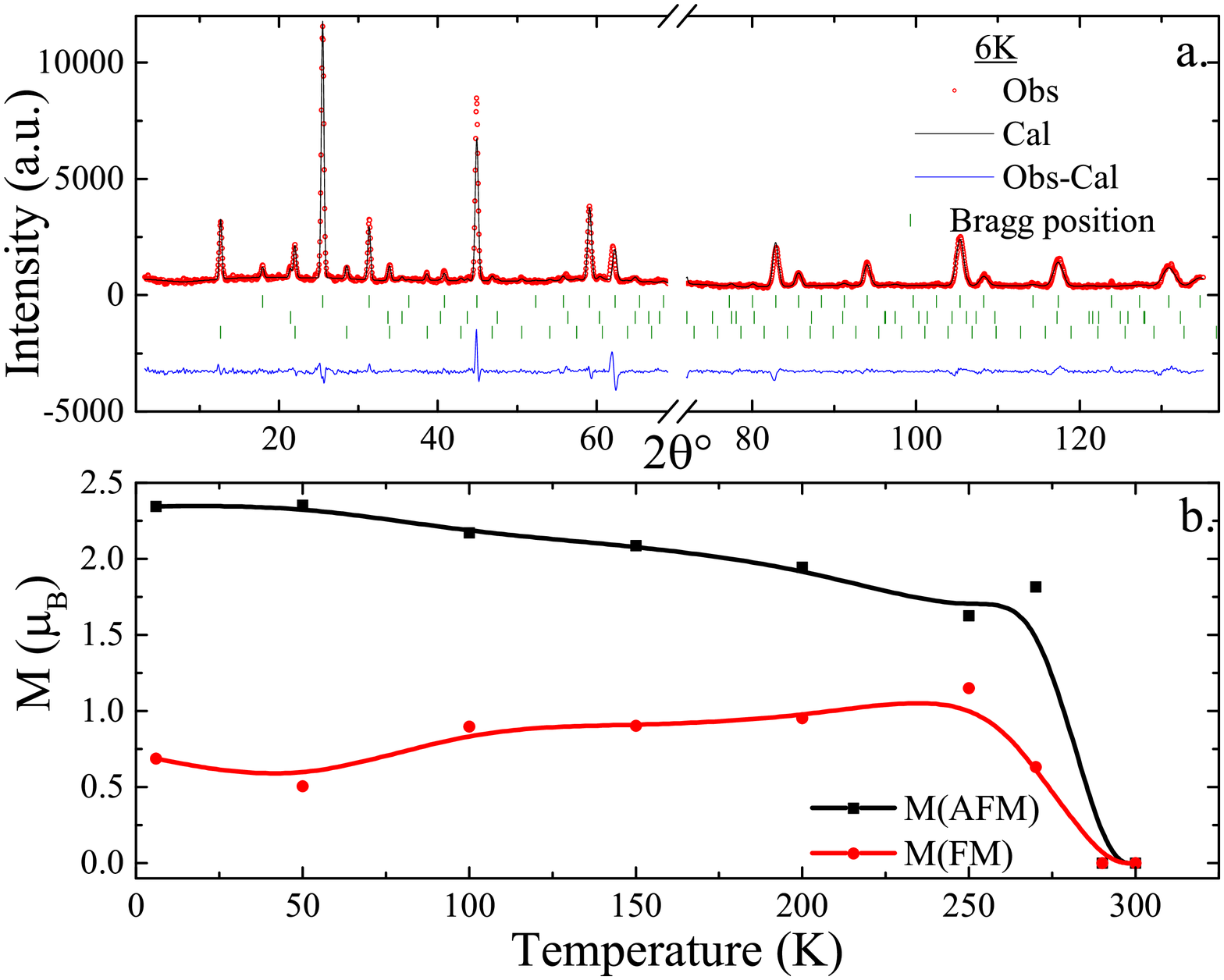}
\caption{a. Rietveld refined neutron diffraction patterns of Mn$_{3}$SnC recorded below the magnetic ordering temperature (T = 6K). b. Variation of refined magnetic moments  of the two species of Mn atoms in Mn$_{3}$SnC magnetic structure as a function of temperature.}
\label{fig:neutron4}
\end{center}
\end{figure}

On cooling below T$_{C} $ in addition to the nuclear scattering peaks defined by the Pm$\overline{3}$m space group superlattice reflections with significant magnetic contriution appear as shown in the diffraction data recorded at 6K in Figure \ref{fig:neutron4}a. These magnetic reflections cannot be indexed on the basis of the obtained chemical structure. They can only be indexed by defining a propagation vector as $k$ = [$\frac{1}{2}$, $\frac{1}{2}$, $0$] and adopting a larger magnetic unit cell of type ($a\sqrt{2}$, $a\sqrt{2}$, $a$) giving rise to a spin alignment shown in Figure \ref{fig:neutron5}. This magnetic structure envisages two magnetic Mn atoms. One, Mn1, has only a FM moment $= 0.7 \pm 0.4 \mu_{B}$ along the 001 axis and the other (Mn2) with only an antiferromagnetic component are arranged in a $\lq$square configuration' in the a-b plane with a net moment $= 2.3 \pm 0.1 \mu_{B}$. Thermal evolution of the magnetic moments of the two species of Mn atoms present in the magnetic structure of Mn$_{3}$SnC is shown in Figure \ref{fig:neutron4}b.  With the lowering of temperature below transition temperature one can see a sharp increase in the magnetic moment of Mn2 species which is as expected for a first order transformation. While the moment of Mn1 which contributes purely to the FM component exhibits a comparatively gradual increase. Furthermore, with lowering of temperature, while the magnetic moment of Mn2 continuously increases, that of Mn1 shows a slight decrease.  The behaviour of Mn magnetic moments near transition temperature, gives an impression that only Mn2 atoms contribute to the first order transformation in Mn$_3$SnC. Eventhough, magnetic structure has a tetragonal symmetry, no structural distortions were visible either in neutron diffraction or in synchrotron XRD measurements at lower temperatures. 

\begin{figure}
\begin{center}
\includegraphics[width=\columnwidth]{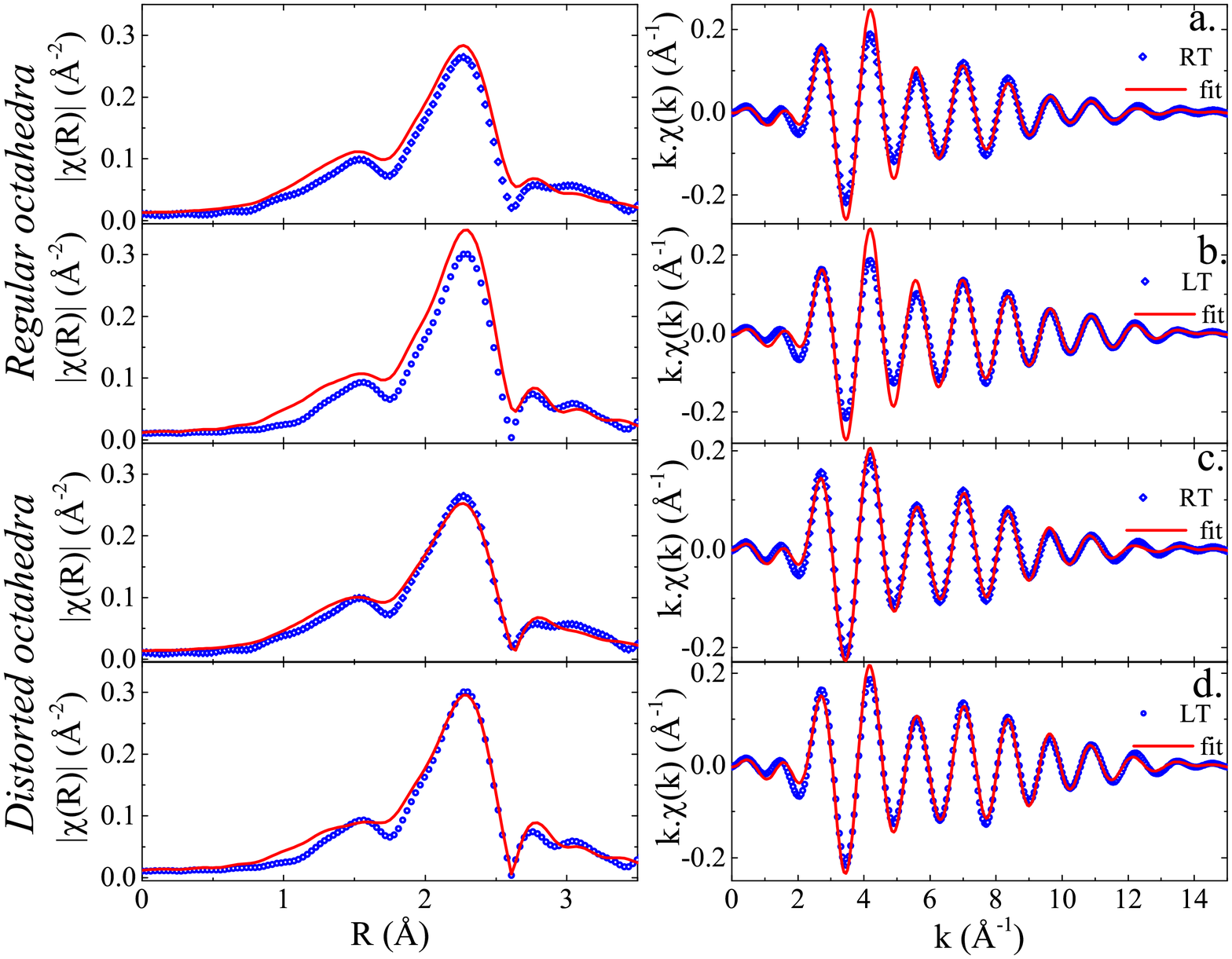}
\caption{Comparison between fitted FT Mn K-edge EXAFS data recorded for Mn$  _{3}$SnC at RT and LT using the two structural models.}
\label{fig:chir}
\end{center}
\end{figure}

To check the possibility of existence of any local structural distortions around Mn in Mn$_3$SnC which could be responsible for the above behaviour of Mn spins, EXAFS data recorded at the Mn K edge at RT and LT was analyzed. For this $\chi(k)$ signal in the range 2\AA$^{-1}$ to 13\AA$^{-1}$ was Fourier transformed (FT) to R space. The magnitude of  FT of EXAFS data at the two temperatures in the R range of 0 to 6 \AA~ are presented in Figure \ref{fig:chir}. The plot shows two distinct peaks between 1\AA~ to 3\AA~ corresponding to Mn-C and equidistant Mn-Mn and Mn-Sn correlations. The scattering contributions from each of these correlations were obtained using FEFF6.01 \cite{Zabinsky199552} and were used in fitting the experimental data in R space in the range 1\AA~ to 3\AA. Since the observed crystal structure at RT is cubic, the data was initially fitted using the structural restrictions imposed by the Pm$\bar3$m space group. These restrictions implied that nearest neighbour Mn-C correlation had a bond distance equal to $a/2$ = 1.998\AA~ and both eight neighboured Mn-Mn and four neighboured Mn-Sn correlations had a bond distance equal to $a/\sqrt2$ = 2.826\AA. The resulting fit in the R space and the back transformed $k$ space is shown in Figure \ref{fig:chir}a. As can be seen, the fit was not quite good (R-factor = 0.04720). An attempt to fit the data by relaxing the above restrictions resulted in a very good fit. However, Mn-Mn bond length was obtained to be slightly shorter than Mn-Sn bond distance. This hints towards the presence of local structural distortions. Taking into account the crystal and magnetic structure obtained from x-ray and neutron diffraction a model was designed consisting of a fraction $\lq x$' of shorter and longer bond distances of Mn with its near neighbours while still preserving the structural restrictions imposed by the crystal structure. Such a model resulted in a very good fit (R-factor = 0.01399),(see Figure \ref{fig:chir}c) wherein about two third of the nearest neighbour Mn-C distances are shorter while the remaining one third are longer.  Likewise around two third Mn-Mn distances and about one third of Mn-Sn distances were found to be shorter and the remaining fraction longer. Such a fitting indicates that Mn$_6$C octahedra distort from their cubic symmetry by elongating along one direction and shrinking along the other two. The structural parameters obtained from the fitting are tabulated in Table \ref{tbl-bonds}. The same model can also be used to fit the LT data equally well but with shorter bonds showing a small increase in their bond lengths and the longer bond distances reducing in their lengths. 

\begin{table*}
\caption{Interatomic bond distance R and mean-square displacement of the path lengths due to thermal or static disorder $\lq \sigma^{2} $' obtained for Mn$  _{3}$SnC compound by fitting Mn K-edge EXAFS to a model representing distorted octahedra. C.N. corresponds to the coordination number, R$_{\rm300K}$ and R$_{\rm100K}$ are interatomic distances calculated from neutron diffraction pattern at 300K and 100K respectively, $\lq x$' denotes the fraction of elongated/shortened bond lengths).}
\label{tbl-bonds}
\begin{tabular}{|c|c|c|c|c|c|c|c|c|c|}
\hline
\multirow{2}{*}{Bond} & \multirow{2}{*}{C.N.} & \multirow{2}{*}{R$_{\rm300K}$(\AA)} & \multirow{2}{*}{$ x $} & \multicolumn{2}{c|}{RT} & \multirow{2}{*}{R$_{\rm100K}$(\AA)} & $ x $ & \multicolumn{2}{c|}{LT} \\ \cline{5-6} \cline{9-10} 
 &  &  &  & R(\AA) & $ \sigma^{2} $(\AA$ ^{2} $) &  &  & R(\AA) & $ \sigma^{2} $(\AA$ ^{2} $) \\ \hline
\multirow{2}{*}{Mn-C} & \multirow{2}{*}{2} & \multirow{2}{*}{1.998} & 0.37(3) & 2.032(2) & \multirow{2}{*}{0.006(3)} & \multirow{2}{*}{1.997} & 0.32(4) & 2.029(3) & \multirow{2}{*}{0.009(5)} \\ \cline{4-5} \cline{8-9}
 &  &  & 0.63(3) & 1.958(2) &  &  & 0.68(4) & 1.961(3) &  \\ \hline
\multirow{2}{*}{Mn-Mn} & \multirow{2}{*}{8} & \multirow{2}{*}{2.826} & 0.37(3) & 2.874(2) & \multirow{2}{*}{0.006(1)} & \multirow{2}{*}{2.824} & 0.32(4) & 2.869(3) & \multirow{2}{*}{0.005(1)} \\ \cline{4-5} \cline{8-9}
 &  &  & 0.63(3) & 2.769(2) &  &  & 0.68(4) & 2.773(3) &  \\ \hline
\multirow{2}{*}{Mn-Sn} & \multirow{2}{*}{4} & \multirow{2}{*}{2.826} & 0.37(3) & 2.769(2) & \multirow{2}{*}{0.008(1)} & \multirow{2}{*}{2.824} & 0.32(4) & 2.773(3) & \multirow{2}{*}{0.006(1)} \\ \cline{4-5} \cline{8-9}
 &  &  & 0.63(3) & 2.874(2) &  &  & 0.68(4) & 2.869(3) &  \\ \hline
\end{tabular}
\end{table*}

Even though no structural disorder is seen in room temperature x-ray or neutron diffraction patterns of Mn$_{3}$SnC, EXAFS analysis indicate a presence of local structural disorder around Mn atoms in the paramagnetic phase. The presence of such a distortion at RT in Mn$_3$SnC could be either due to the proximity to the magnetostructural transition temperature or due to the presence of larger atom like Sn. It may be mentioned here that such local structural distortions have not been hitherto reported in other antiperovskite compounds above their transformation temperature.  Larger atoms like Sn could induce local strains which may be alleviated by a structural distortions while maintaining the cubic symmetry of the overall crystal structure. EXAFS studies also show that though distortions persist even in the magnetically ordered phase the shorter bond distances tend to elongate while the longer bond distances shrink. 

The presence of such a disorder could also be the reason behind the change in magnetic propogation vector from (0.5, 0.5, 0.5) in Mn$_3$GaC to (0.5, 0.5, 0) in Mn$_3$SnC. It is known that in antiperovskites, the nature of magnetic order critically depends on the competition between the interaction strengths of Mn$-$Mn nearest neighbour $J_1$ and next nearest neighbour $J_2$ interactions which are of opposite sign. The frustration caused due to this competition is often aliviated by a structural transformation. In the case of Mn$_3$SnC, the presence of a local structural distortions could be the reason for its unusual magnetic groud state.

\section{Conclusions}
A systematic exploration of the Mn$_3$SnC crystal structure using a combination of x-ray and neutron diffraction and x-ray absorption fine structure (XAFS) spectroscopy was carried out to understand the nature of the first order magnetic transformation in the compound. Although x-ray and neutron diffraction patterns show no structural distortions in the compound at low temperatures, they indicate a cubic-cubic volume discontinuous transition accompanying the magnetic transition (T$\sim$279K). However EXAFS results suggest a structural distortion of the Mn$_6$C octahedra from their cubic symmetry due to presence of larger atoms of Sn. These distortions critically affect the magnetic order of Mn$_3$SnC ground state consisting of antiferromagnetic and ferromagnetic Mn atoms.

\section*{Acknowledgments}
This work is supported by Board of Research in Nuclear Sciences (BRNS) under the project 2011/37P/06. Further financial support and experimental facilitation granted by Department of Science and Technology, India and Saha Institute of Nuclear Physics, India at the Indian Beamline, Photon Factory, KEK, Japan is gratefully acknowledged. M/s Devendra D. Buddhikot, Ganesh Jangam and Dr. V. Srihari are acknowledged for the experimental assistance.

\bibliographystyle{elsarticle-num}


\bibliography{references}

\begin{thebibliography}{10}
\expandafter\ifx\csname url\endcsname\relax
  \def\url#1{\texttt{#1}}\fi
\expandafter\ifx\csname urlprefix\endcsname\relax\def\urlprefix{URL }\fi
\expandafter\ifx\csname href\endcsname\relax
  \def\href#1#2{#2} \def\path#1{#1}\fi

\bibitem{He2001411}
T.~He, Q.~Huang, A.~Ramirez, Y.~Wang, K.~Regan, N.~Rogado, M.~Hayward, M.~Haas,
  J.~Slusky, K.~Inumara, et~al., Nature 411~(6833) (2001) 54--56.

\bibitem{Uehara200776}
M.~Uehara, T.~Yamazaki, T.~Kôri, T.~Kashida, Y.~Kimishima, I.~Hase, Journal of
  the Physical Society of Japan 76~(3) (2007) 034714.

\bibitem{Kamishima200063}
K.~Kamishima, T.~Goto, H.~Nakagawa, N.~Miura, M.~Ohashi, N.~Mori, T.~Sasaki,
  T.~Kanomata, Phys. Rev. B 63 (2000) 024426.

\bibitem{Asano200892}
K.~Asano, K.~Koyama, K.~Takenaka, Applied Physics Letters 92~(16) (2008)
  161909.

\bibitem{Takenaka200587}
K.~Takenaka, H.~Takagi, Applied Physics Letters 87~(26) (2005) 261902.

\bibitem{Takenaka200892}
K.~Takenaka, K.~Asano, M.~Misawa, H.~Takagi, Applied Physics Letters 92~(1)
  (2008) 011927.

\bibitem{Iikubo2008101}
S.~Iikubo, K.~Kodama, K.~Takenaka, H.~Takagi, M.~Takigawa, S.~Shamoto, Phys.
  Rev. Lett. 101 (2008) 205901.

\bibitem{Huang200893}
R.~Huang, L.~Li, F.~Cai, X.~Xu, L.~Qian, Applied Physics Letters 93~(8) (2008)
  081902.

\bibitem{Takenaka200994}
K.~Takenaka, H.~Takagi, Applied Physics Letters 94~(13) (2009) 131904.

\bibitem{Wang200985}
B.~S. Wang, P.~Tong, Y.~P. Sun, X.~Luo, X.~B. Zhu, G.~Li, X.~D. Zhu, S.~B.
  Zhang, Z.~R. Yang, W.~H. Song, J.~M. Dai, EPL 85~(4) (2009) 47004.

\bibitem{Howe19572}
L.~Howe, H.~P. Myers, Philosophical Magazine 2~(16) (1957) 554--560.

\bibitem{Lorthioir19738}
G.~Lorthioir, M.~E. Fruchart, M.~Nardin, P.~l'Héritier, R.~Fruchart, Materials
  Research Bulletin 8~(9) (1973) 1027--1034.

\bibitem{Heritier197712}
P.~l'Heritier, J.~P. Senateur, R.~Fruchart, D.~Fruchart, E.~F. Bertaut, Mater
  Res Bull 12~(5) (1977) 533--541.

\bibitem{Li200572}
Y.~B. Li, W.~F. Li, W.~J. Feng, Y.~Q. Zhang, Z.~D. Zhang, Phys. Rev. B 72
  (2005) 024411.

\bibitem{Fruchart19708}
D.~Fruchart, E.~Bertaut, F.~Sayetat, M.~N. Eddine, R.~Fruchart, J.~Sénateur,
  Solid State Commun 8~(2) (1970) 91--99.

\bibitem{Lewis200393}
M.-H. Yu, L.~H. Lewis, A.~R. Moodenbaugh, J. Appl. Phys. 93~(12) (2003)
  10128--10130.

\bibitem{Carvajal1993192}
J.~Rodríguez-Carvajal, Physica B: Condensed Matter 192~(1–2) (1993) 55--69.

\bibitem{Giuliana2009190}
A.~D. Cicco, G.~Aquilanti, M.~Minicucci, E.~Principi, N.~Novello, A.~Cognigni,
  L.~Olivi, Journal of Physics: Conference Series 190~(1) (2009) 012043.

\bibitem{Ravel200512}
B.~Ravel, M.~Newville, Journal of Synchrotron Radiation 12~(4) (2005) 537--541.

\bibitem{Zabinsky199552}
S.~I. Zabinsky, J.~J. Rehr, A.~Ankudinov, R.~C. Albers, M.~J. Eller, Phys. Rev.
  B 52 (1995) 2995--3009.

\end{thebibliography}

\end{document}